\documentclass{aa}

\usepackage{graphicx}
\usepackage{txfonts}

\newcommand{\myemail}{mj.michalowski@gmail.com}
\newcommand{\urltt}[1]{\url{\texttt{#1}}}

\newcommand{\msun}{\mbox{$M_\odot$}}
\newcommand{\msunyr}{\mbox{\msun\,yr$^{-1}$}}
\newcommand{\hi}{\sc Hi}
\newcommand{\mhi}{M_{\rm HI}}
\newcommand{\mhtwo}{M_{\rm H2}}
\newcommand{\kms}{\mbox{km\,s$^{-1}$}}
\newcommand{\jykms}{\mbox{Jy\,km\,s$^{-1}$}}

\newcommand{\ngc}{NGC\,2770}
\urldef{\hyperleda}\url{http://leda.univ-lyon1.fr/ledacat.cgi?ngc%202770}

\begin{document}

\title{NGC\,2770: High supernova rate due to interaction
}

\titlerunning{NGC\,2770}
\authorrunning{Micha{\l}owski et al.}
 
\author{Micha{\l}~J.~Micha{\l}owski\inst{\ref{inst:uam},\ref{inst:cal},\ref{fulb}}, 
    Christina Th\"one\inst{\ref{inst:iaa}},
    Antonio de Ugarte Postigo\inst{\ref{inst:iaa},\ref{inst:dark}},
    Jens~Hjorth\inst{\ref{inst:dark}},
    \newline Aleksandra Le\'sniewska\inst{\ref{inst:uam}},
        Natalia~Gotkiewicz\inst{\ref{inst:umk}},
        Wojciech Dimitrov\inst{\ref{inst:uam}},
        Maciej P.~Koprowski\inst{\ref{inst:umk}},
        Peter Kamphuis\inst{\ref{inst:airub}}
        }

\institute{
Astronomical Observatory Institute, Faculty of Physics, Adam Mickiewicz University, ul.~S{\l}oneczna 36, 60-286 Pozna{\'n}, Poland \myemail  \label{inst:uam}
\and
TAPIR, Mailcode 350-17, California Institute of Technology, Pasadena, CA 91125, USA  \label{inst:cal}
\and
Fulbright Senior Award Fellow \label{fulb}
\and
Instituto de Astrof\' isica de Andaluc\' ia (IAA-CSIC), Glorieta de la Astronom\' ia, s/n, E-18008, Granada, Spain \label{inst:iaa}
\and
DARK, Niels Bohr Institute, University of Copenhagen, Lyngbyvej 2, 2100 Copenhagen, Denmark\label{inst:dark}
\and
Institute of Physics, Faculty of Physics, Astronomy and Informatics, Nicolaus Copernicus University, Grudzi\c{a}dzka 5, 87-100 Toru\'{n}, Poland\label{inst:umk}
\and
Ruhr-Universit\"at Bochum, Faculty of Physics and Astronomy, Astronomical Institute, 44780 Bochum, Germany\label{inst:airub}
}


\abstract
{Galaxies that hosted many core-collapse supernova (SN) explosions can be used to study the conditions necessary for the formation of massive stars.
{\ngc} was dubbed an SN factory because it hosted four core-collapse SNe in 20 years (three type Ib and one type IIn).
Its  star formation rate (SFR) was reported to not be enhanced and, therefore, not compatible with such a high SN rate.
}
{We aim to explain the high SN rate of {\ngc}.}
{We used archival {\hi} line data for {\ngc} 
and reinterpreted  the H$\alpha$ and optical continuum data.}
%
{Even though the continuum-based SFR indicators do not yield high values, the dust-corrected H$\alpha$ luminosity implies a high SFR, consistent with the high SN rate. Such a disparity between the SFR estimators is an indication of recently enhanced star formation activity because the continuum indicators trace long timescales of the order of 100\,Myr, unlike the line indicators, which trace timescales of the order of 10\,Myr.
Hence, the unique feature of {\ngc} compared to other galaxies is the fact that it was observed very shortly after the enhancement of the SFR. It also has high dust extinction, $E(B-V)$ above 1 mag.
We provide support for the hypothesis that the increased SFR in {\ngc} is due to the interaction with its companion galaxies. We report an {\hi} bridge between {\ngc} and its closest companion and the existence of a total of four companions within 100\,kpc (one identified for the first time).
There are no clear {\hi} concentrations close to the positions of SNe in {\ngc} such as those detected for hosts of gamma-ray bursts (GRBs) and broad-lined SNe type Ic (IcBL). This suggests that the progenitors of type Ib SNe are not born out of recently accreted atomic gas, as was suggested for GRB and IcBL SN progenitors.
}
%
{}


\maketitle

\section{Introduction}

\newlength{\szerkol}

\setlength{\szerkol}{0.4\textwidth}

\begin{figure*}
\begin{center}
\begin{tabular}{|c|c|}
\hline
\includegraphics[width=\szerkol,clip]{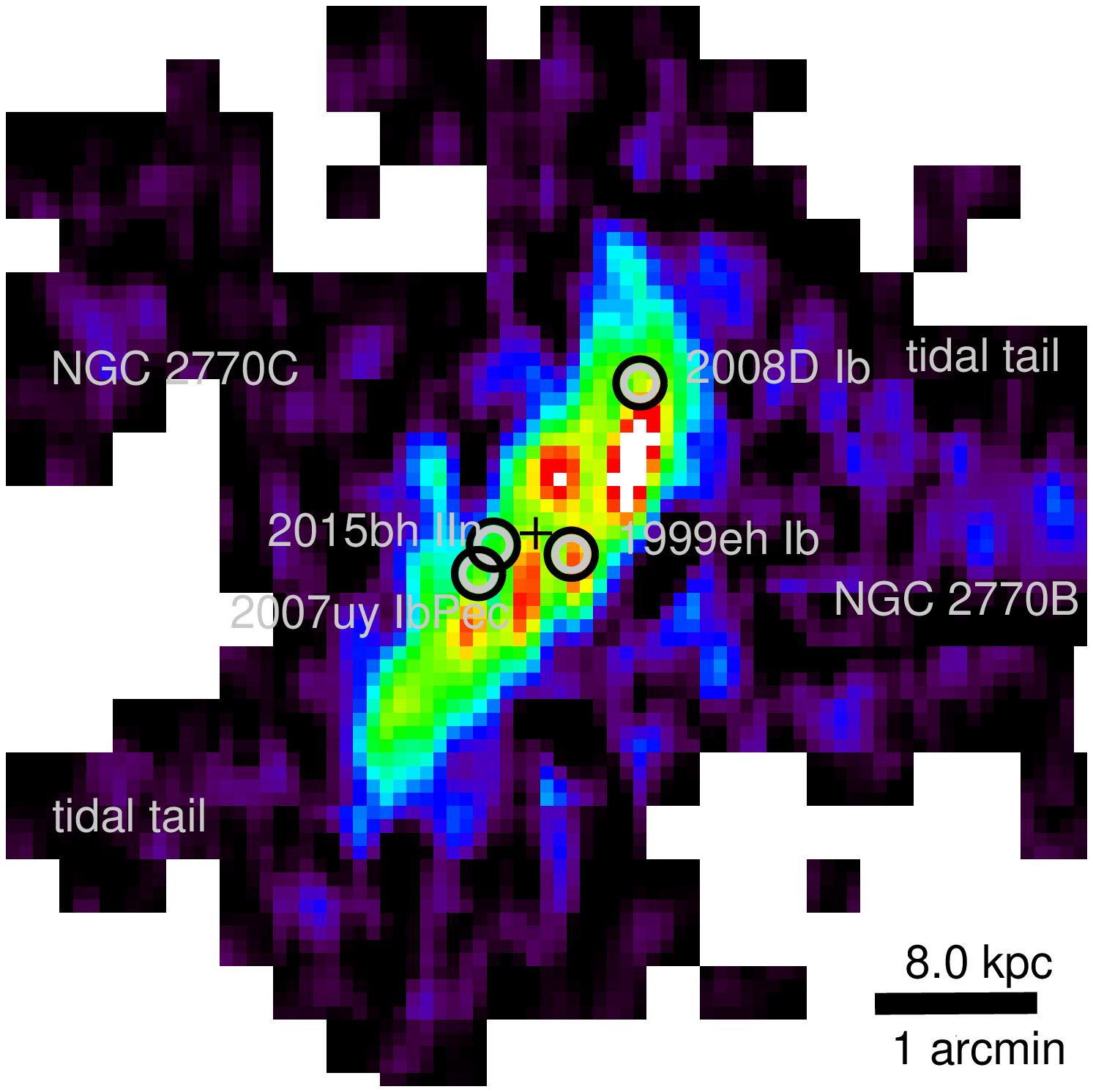} & 
\includegraphics[width=\szerkol,clip]{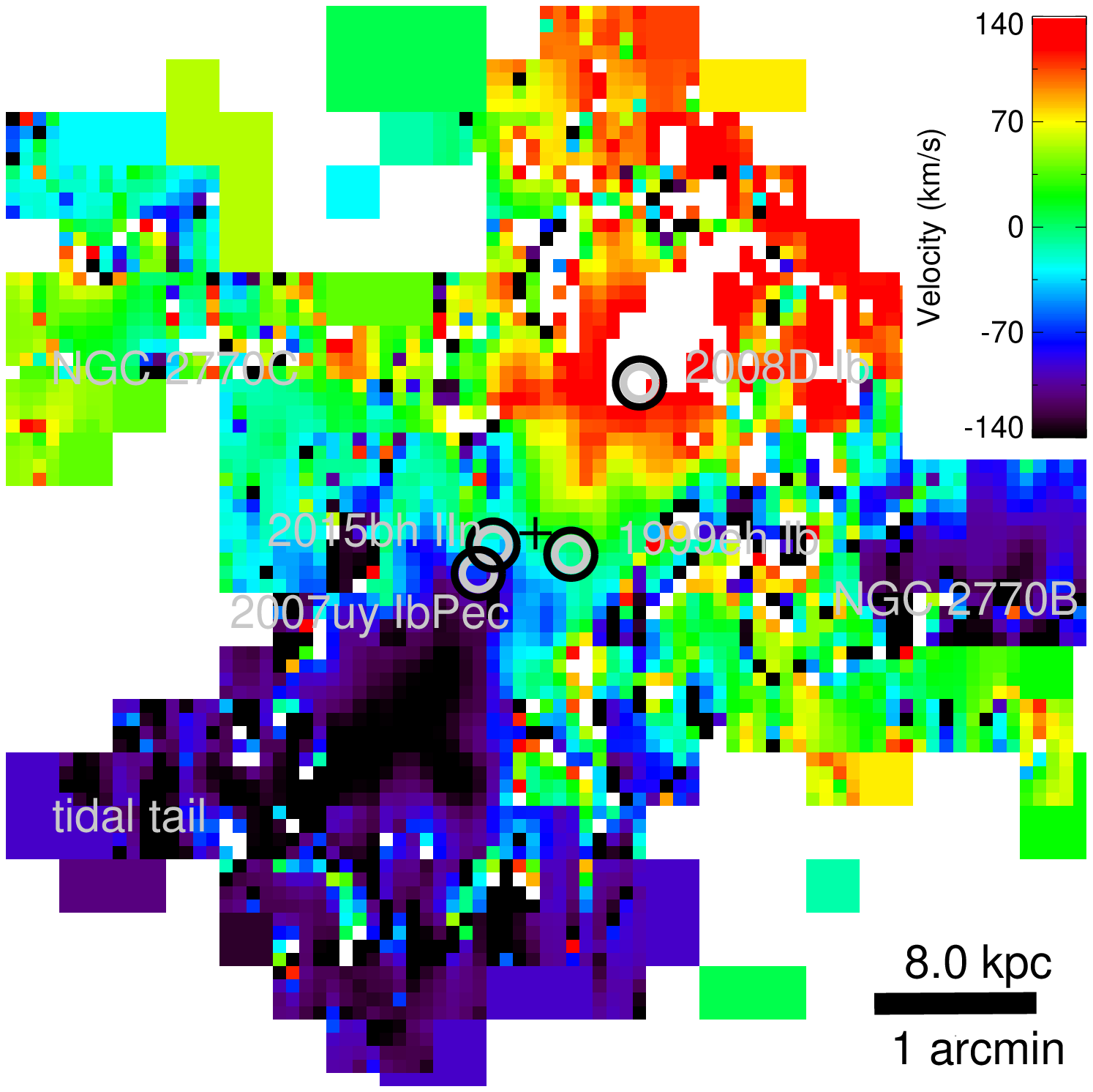}  \\
\hline
\includegraphics[width=\szerkol,clip]{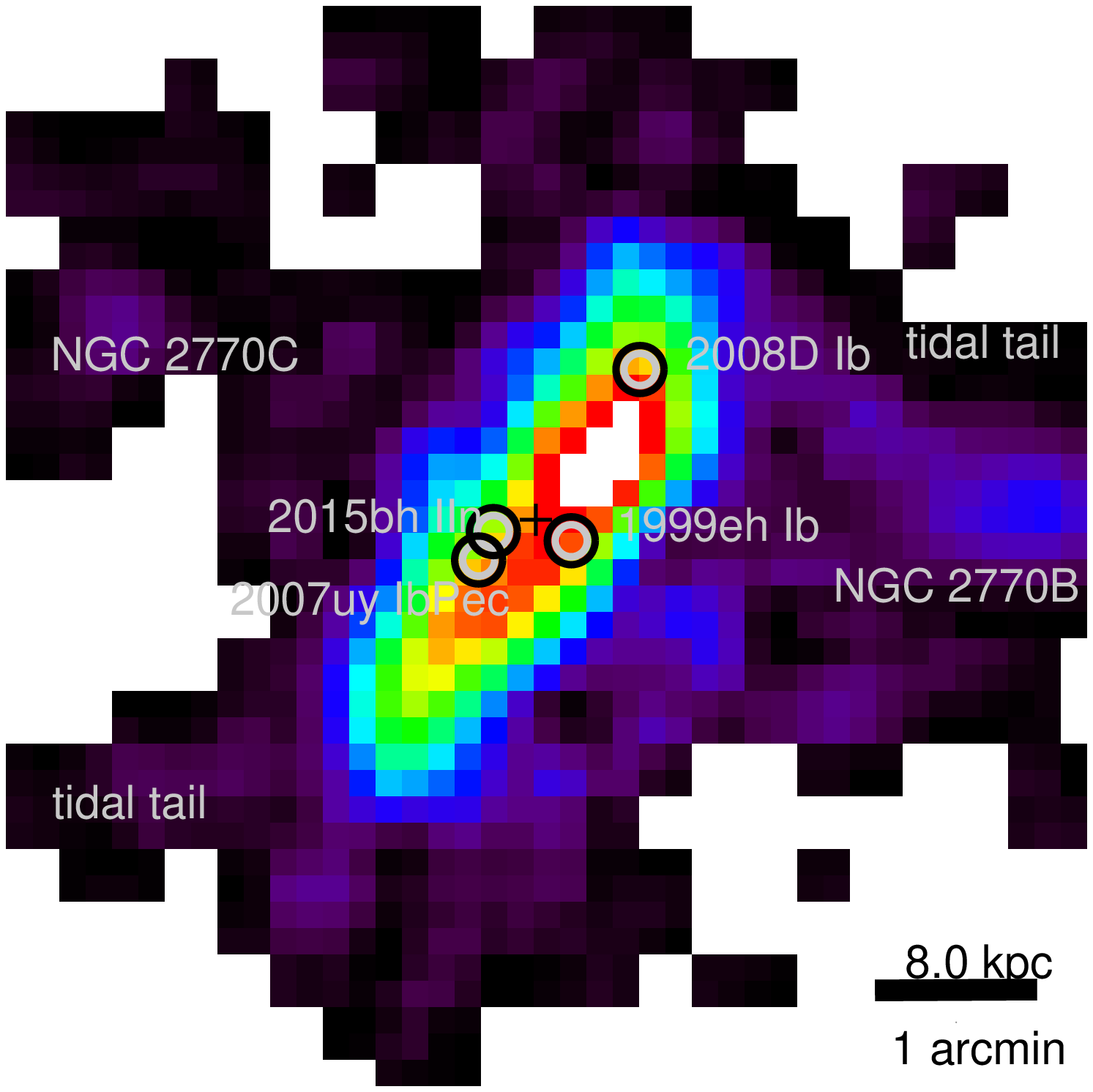} & 
\includegraphics[width=\szerkol,clip]{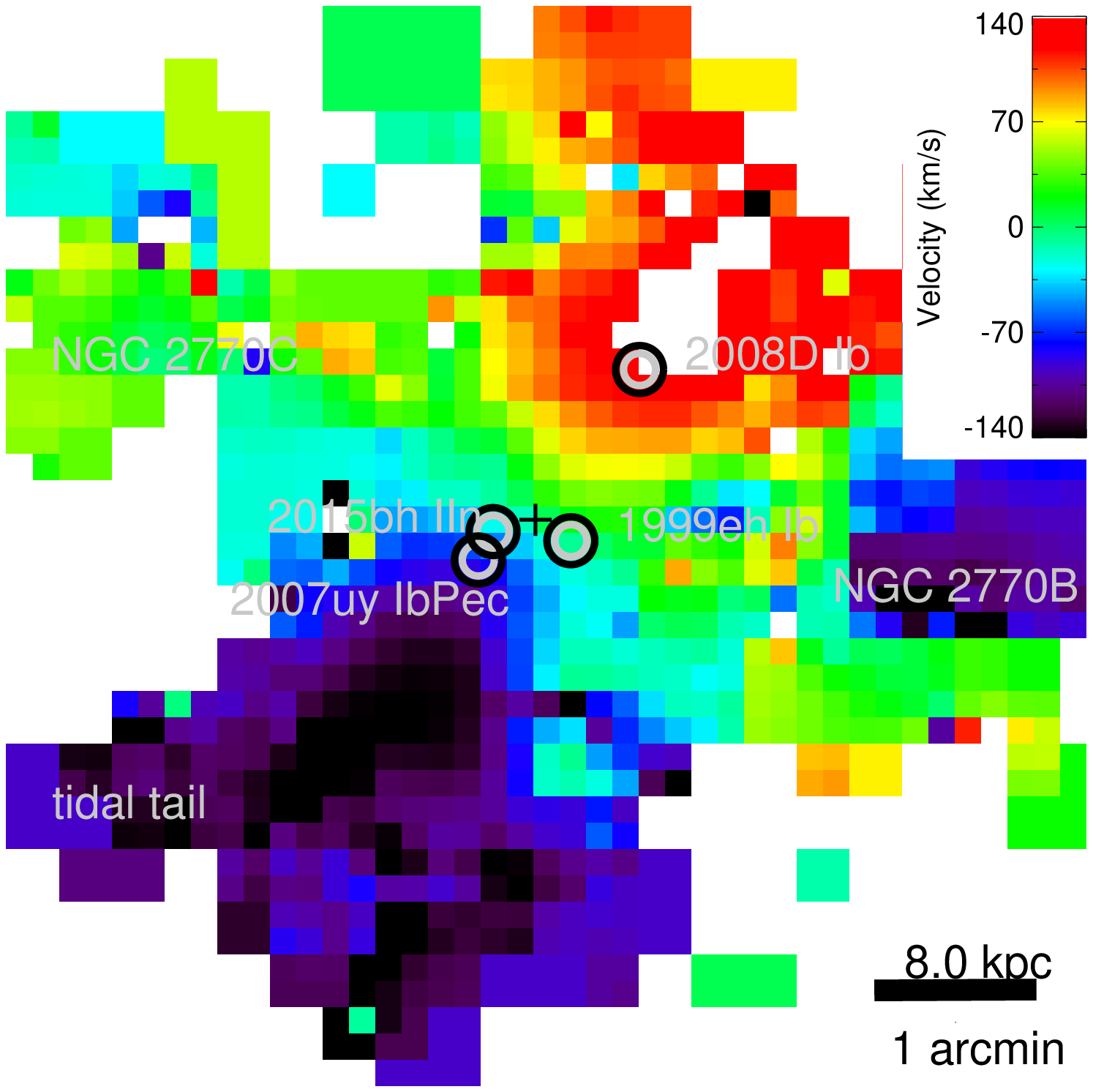}  \\
\hline
\includegraphics[width=\szerkol,clip]{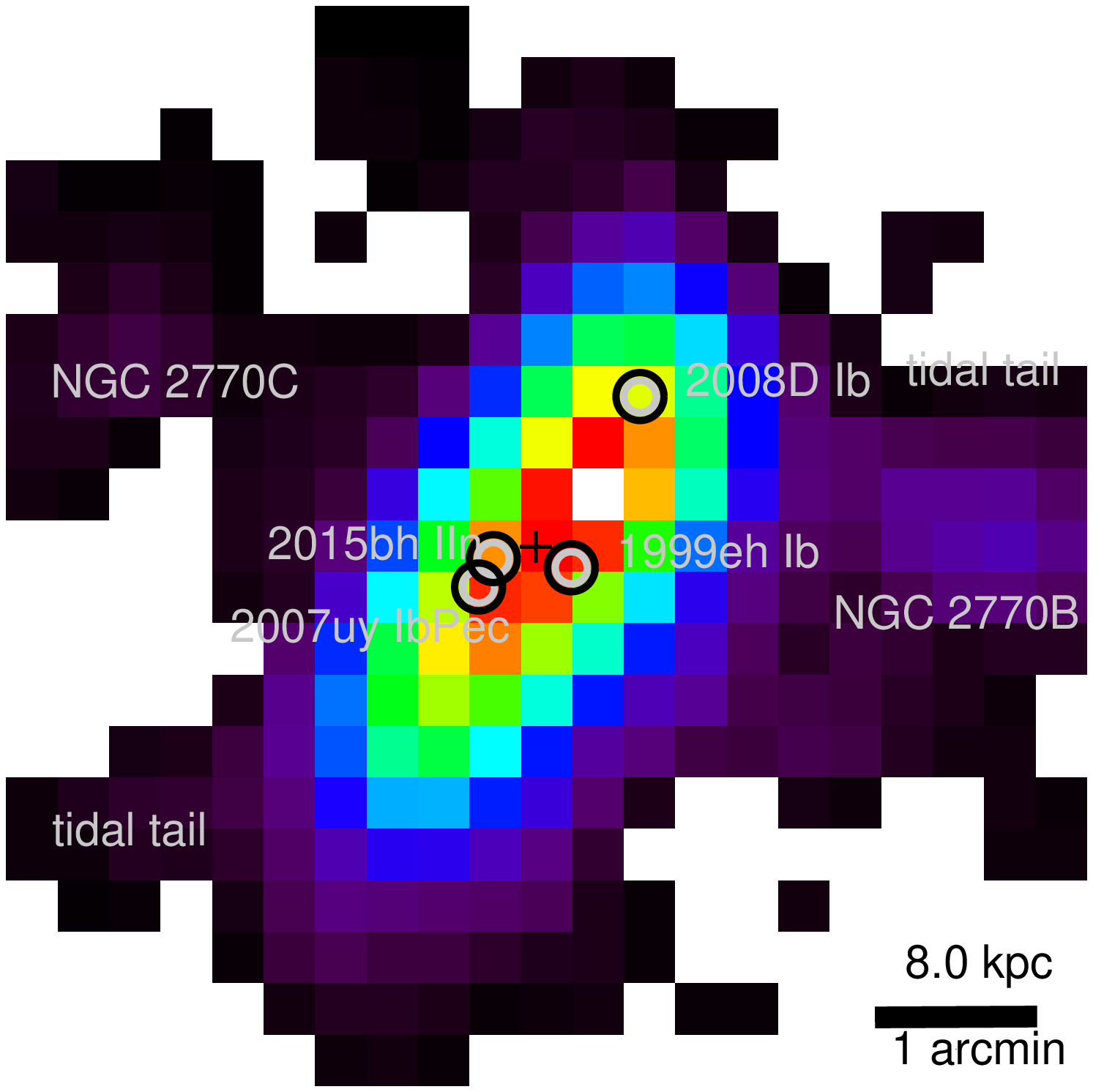} & 
\includegraphics[width=\szerkol,clip]{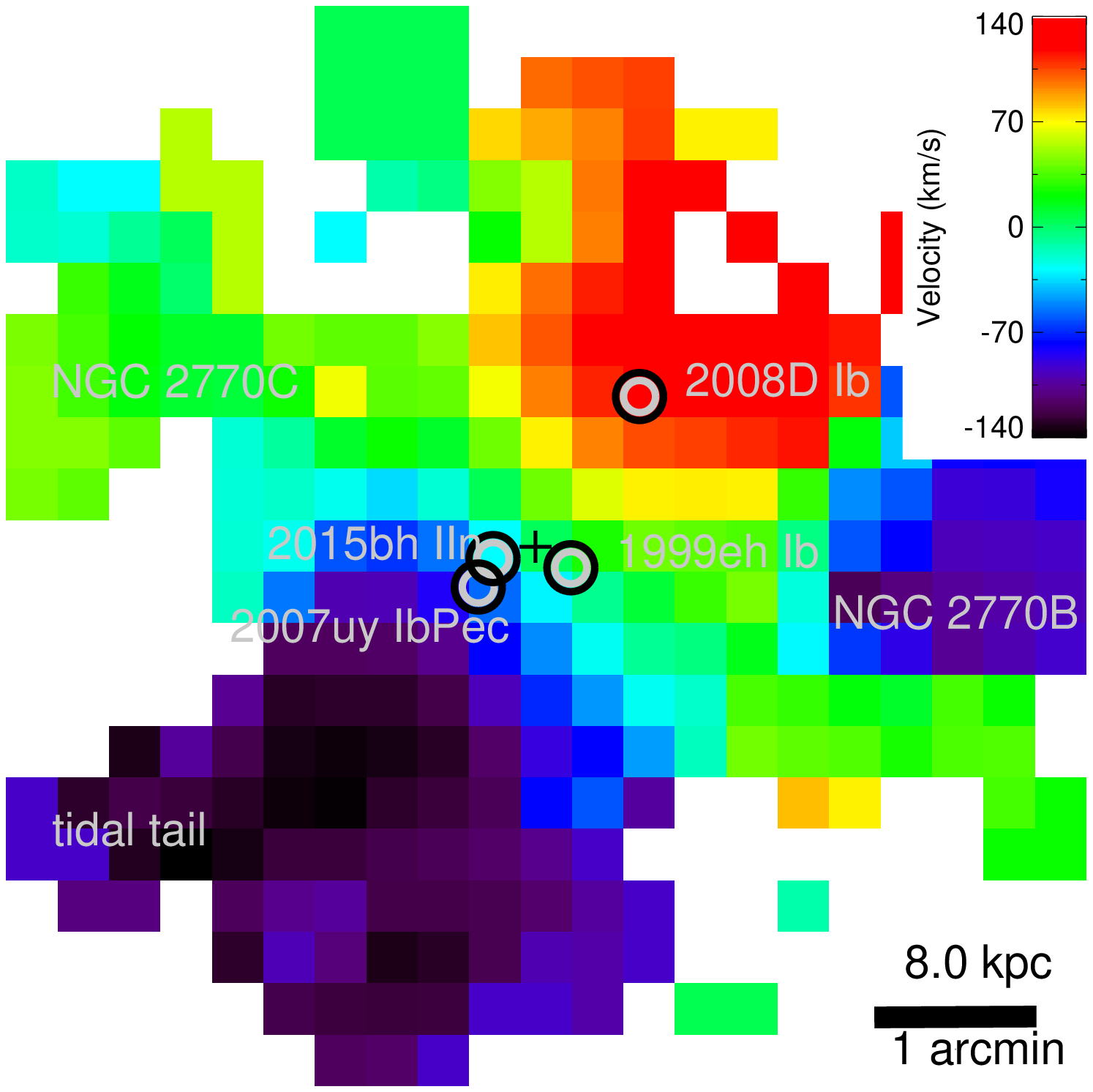} \\ 
\hline
\end{tabular}
\end{center}
\caption{Atomic gas distribution in {\ngc}. The data are at resolutions of 
$24\arcsec\times13\arcsec$ (top), 
$36\arcsec\times30\arcsec$ (middle), and $65\arcsec\times56\arcsec$  (bottom).
Left: zeroth moment maps (integrated emission).
Right:
first moment maps (velocity fields) relative to $z=0.00649$ (1945.65\,{\kms}).
The positions of SNe are marked by grey circles, whereas the black cross shows the position of the optical centre of the galaxy.
Each panel is 6.7{\arcmin} per side. North is up and east is to the left. 
{\ngc}B is at the right edge of the panels and connected to {\ngc} by a tidal feature.
{\ngc}C is at the left edge of the panels.
}
\label{fig:zoom}
\end{figure*}

\begin{figure*}
\begin{center}
\begin{tabular}{|c|c|}
\hline
\includegraphics[width=\szerkol,clip]{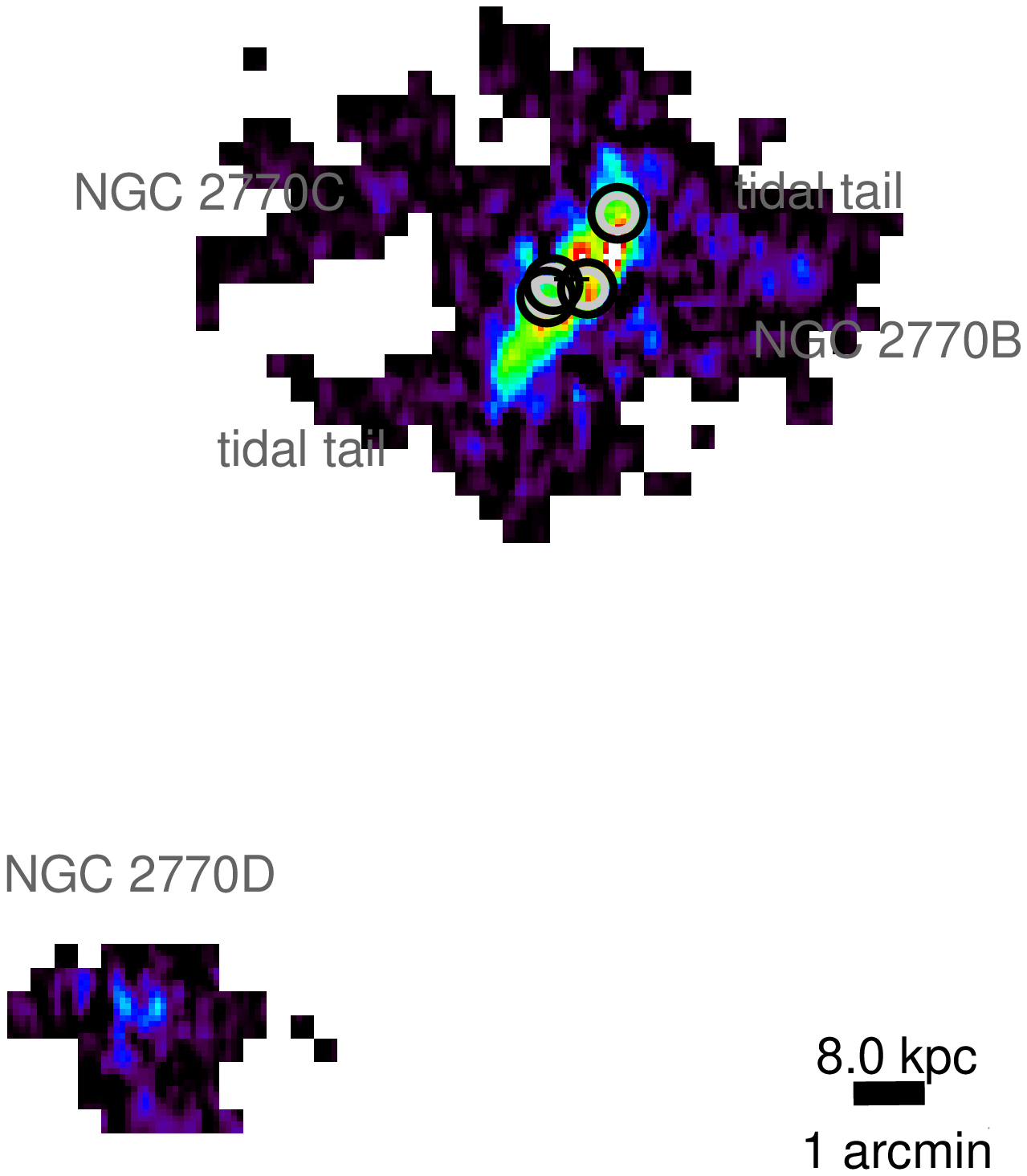} & 
\includegraphics[width=\szerkol,clip]{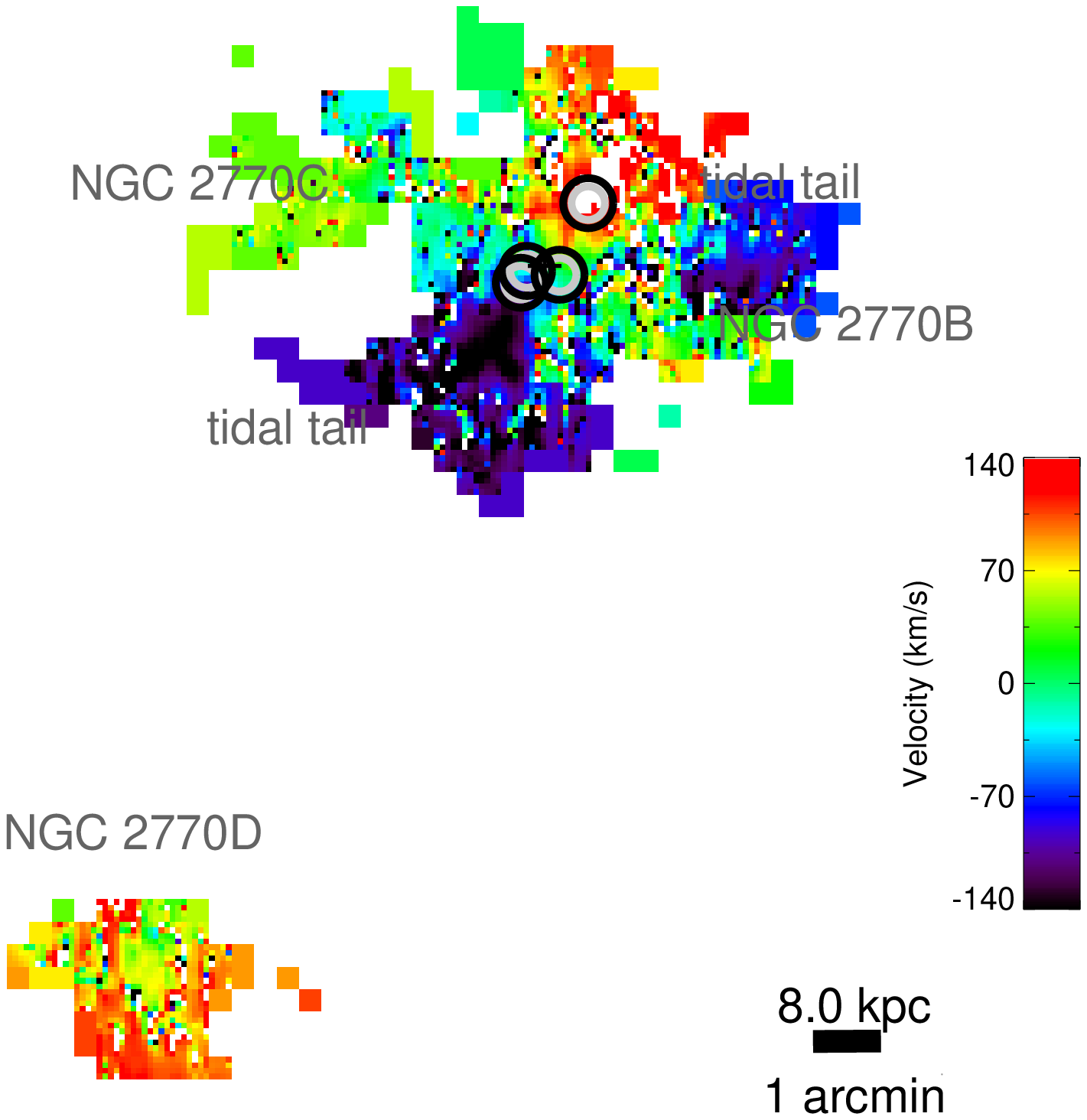}  \\
\hline
\includegraphics[width=\szerkol,clip]{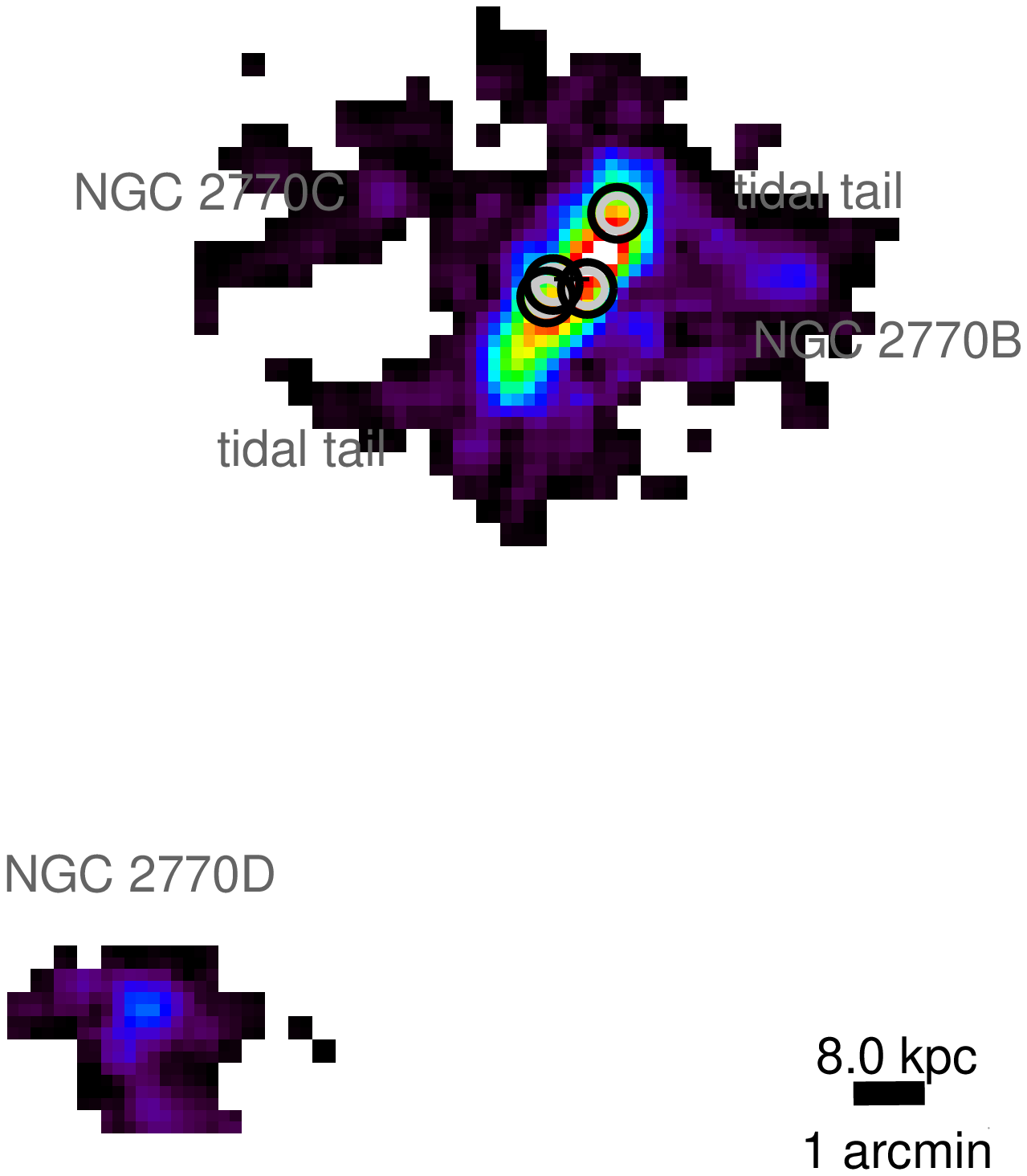} & 
\includegraphics[width=\szerkol,clip]{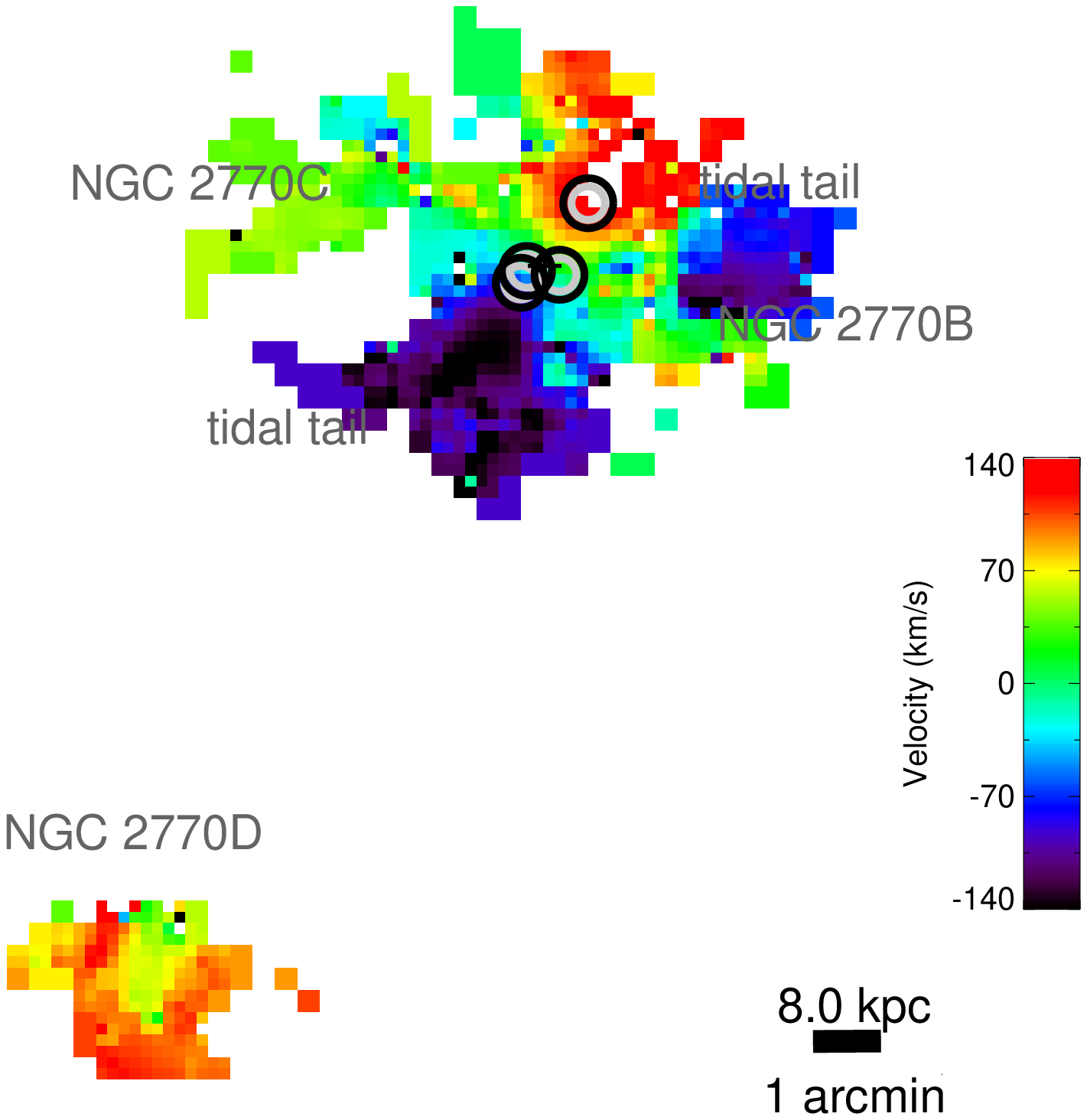}  \\
\hline
\includegraphics[width=\szerkol,clip]{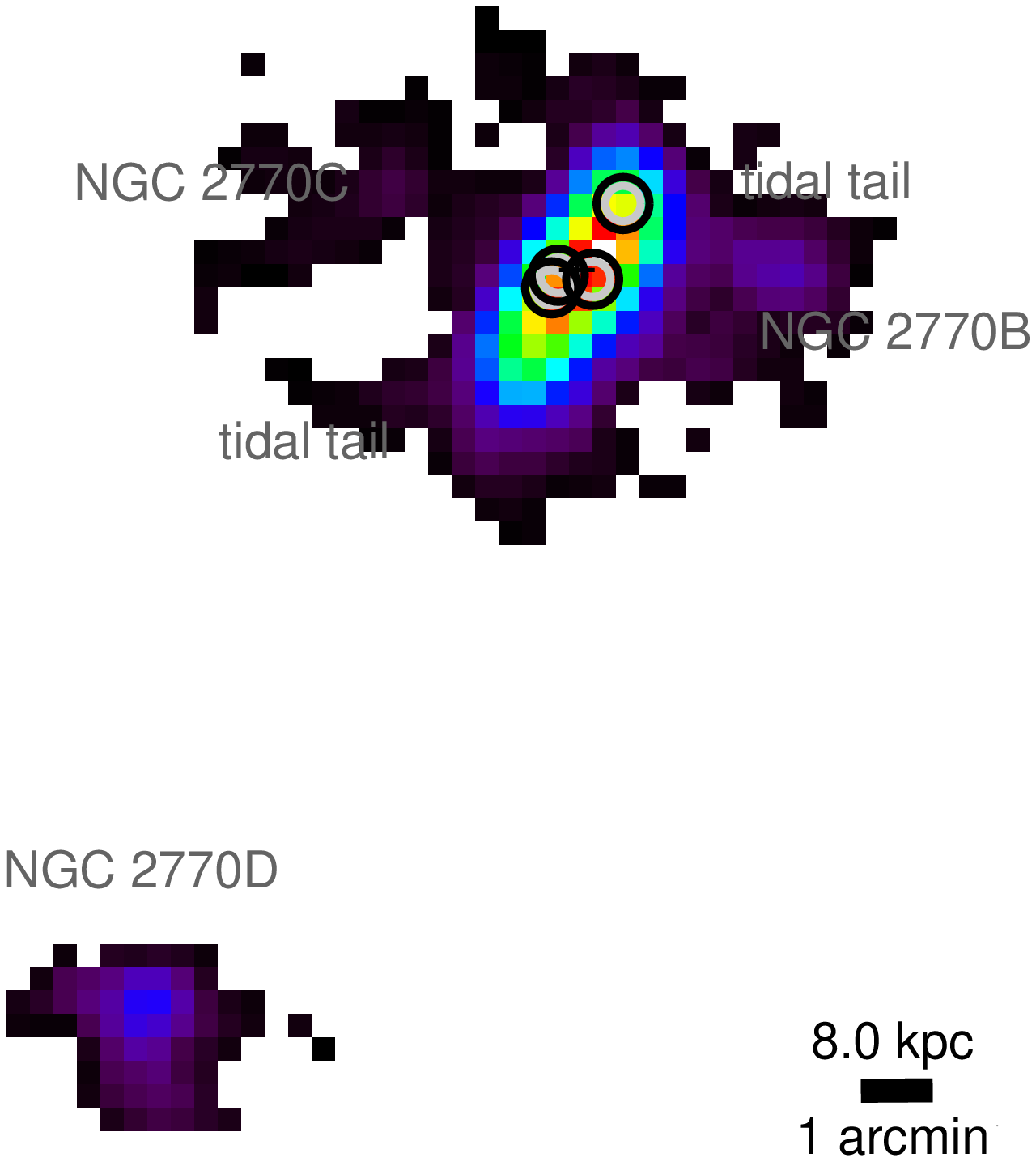} & 
\includegraphics[width=\szerkol,clip]{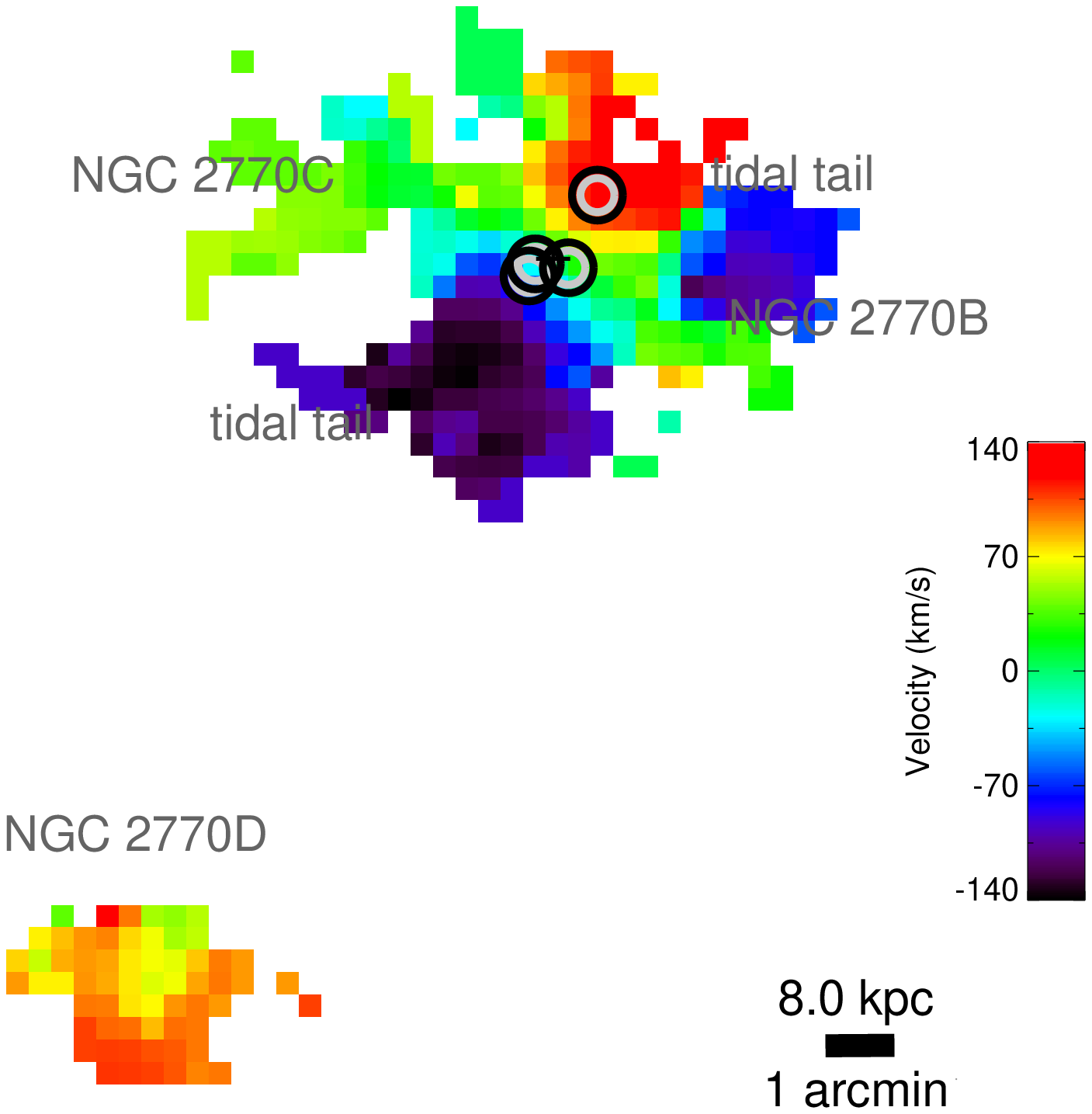} \\ 
\hline
\end{tabular}
\end{center}
\caption{As in Fig.~\ref{fig:zoom}, but zoomed out to show the environment of {\ngc}.
Each panel is 16.7{\arcmin} per side. {\ngc}B and C are $\sim3\arcmin$ (24\,kpc) to the west and east from the centre of {\ngc}, respectively, whereas {\ngc}D is $\sim12\arcmin$ (96\,kpc) to the south-east. There are tidal features extending from both endpoints of the {\ngc} disc.
{\ngc}E is outside the panels to the east of {\ngc}.}
\label{fig:image}
\end{figure*}

\begin{figure*}
\begin{center}
\includegraphics[height=\textwidth,clip,angle=90,viewport=72 0 523 842]{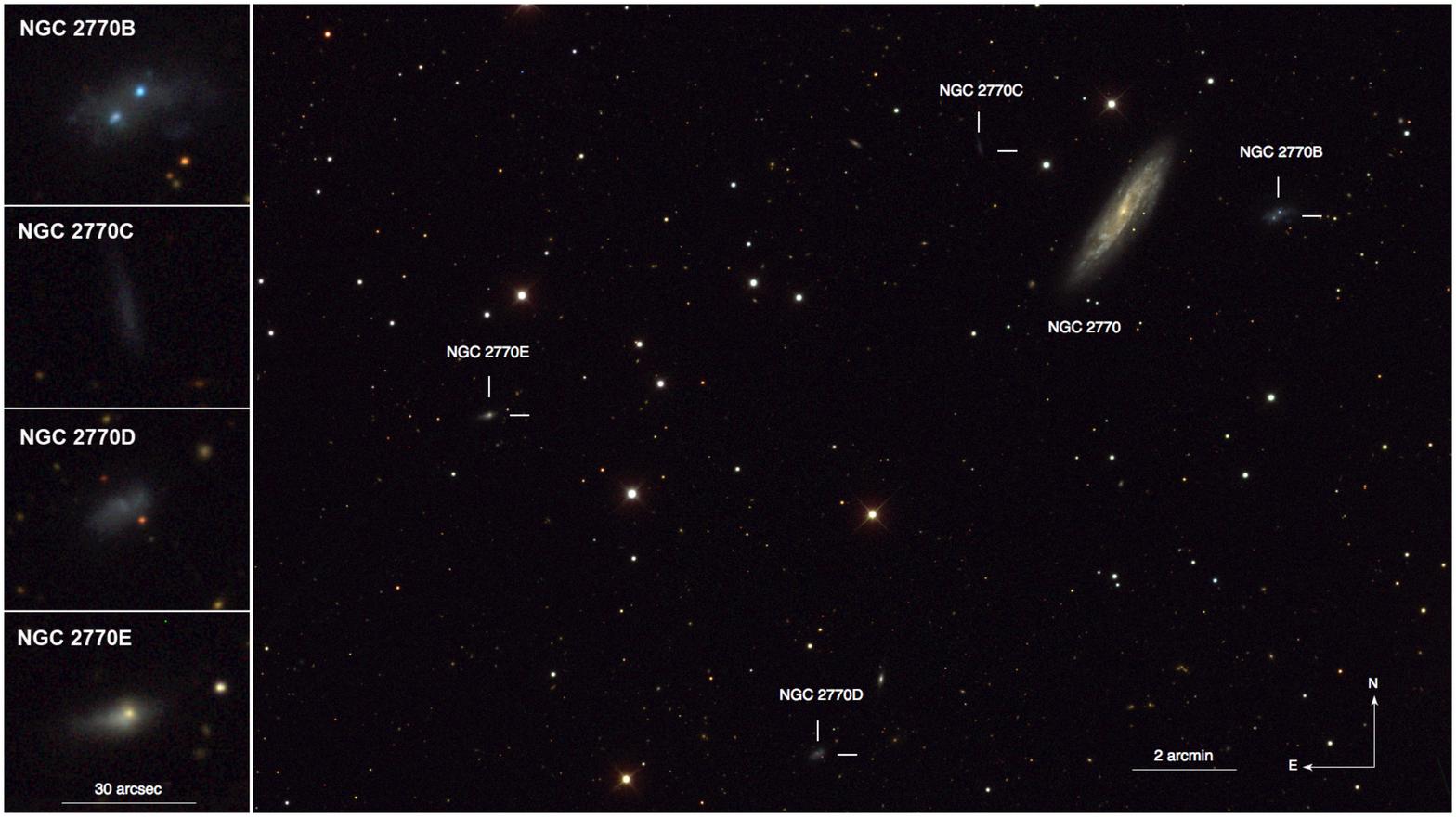} 
\end{center}
\caption{Optical colour image of the NGC 2770 group with all detected companions. For the figure we use $g'$ (blue), $r'$ (yellow), and $i'$ (red) filters from the SDSS DR 16 in a mosaic of 13 plates. On the left we show zoomed in images of the four companions. The different scales are indicated in the images. 
}
\label{fig:opt}
\end{figure*}

Due to their very short lifetimes, the properties and numbers of massive stars in a galaxy can be used to infer the details of star formation. Individual stars can only be detected in the most nearby galaxies, so massive stars in most galaxies can only be studied when they explode as core-collapse supernovae (SNe). 
In the context of using SNe to study star formation, it is especially promising to study galaxies with multiple SNe, as their high SN rates should be connected with conditions necessary for a given type of stellar explosion.
For example, \citet{li95} showed that the number of galaxies hosting multiple SNe does not deviate significantly from a Poissonian distribution, so the hosts of multiple SNe are not inherently different from other galaxies; they just have higher star formation rates (SFRs).
Moreover, \citet{anderson13} find that galaxies with multiple SNe are preferentially of later morphological types, pointing to the likely connection with young stellar populations.

\object{NGC 2770} (\object{UGC 4806}) was dubbed an SN Ib factory \citep{thone09} because it hosted three type Ib SNe in ten years.
Its  SFR was reported to not be enhanced and, therefore, not compatible with such a high SN rate \citep{thone09}.
In 2015, a fourth SN (type IIn) exploded, bringing the current rate to four over the last 20 years.

\object{SN\,1999eh} was the first SN discovered in {\ngc} \citep{hurst99,armstrong00circ} and was classified as type Ib \citep{jha99circ}.
Similarly, \object{SN\,2007uy} was classified as a type Ib \citep{vanderhorst11,roy13} peculiar \citep{modjaz14} SN.
The discovery of \object{SN\,2008D} (also classified as type Ib) in the early phase during the observations of SN\,2007uy allowed the detection of signatures of SN shock breakout \citep{mazzali08,soderberg08,chevalier08,bietenholz09,malesani09,maund09b,modjaz09,tanaka09,gorosabel10,couch11}. 
\object{SN\,2015bh}  was a type IIn SN or a significant outburst of a luminous blue variable (LBV) star. It was proceeded by 20 years of outbursting activity \citep{eliasrosa16,ofek16,thone17,boian18}. 

In the single-star progenitor channel, SNe of both types Ib/c \citep{woosley02} and IIn \citep{kiewe12} are believed to be connected with the endpoints of the evolution of the most massive stars (of the order of $>25\,\msun$),
likely Wolf-Rayet stars for the type Ib/c \citep{gaskell86}. On the other hand, the double-star channel for SNe type Ib/c involves a less massive star ($8$--$20\,\msun$) with the hydrogen envelope removed through interaction in a binary system \citep{podsiadlowski92}. This double-star channel was also claimed for type IIb SNe, that is to say~those transitioning from type II at early times with  hydrogen lines to type Ib at later times with no hydrogen lines  \citep{woosley87,filippenko88,filippenko93,nomoto93,krause08,chevalier10}.

There are a few tens of galaxies hosting multiple SNe, but only one other
galaxy has hosted more than two SNe Ib/c: \object{Arp 299}, the interacting system of \object{NGC 3690,} and \object{IC 694}. This system hosted ten SNe (\citealt{anderson11} and the updated list at the Open Supernova Catalog\footnote{{\tt https://sne.space}}; \citealt{snespace}),  three of which were classified as type Ib (1998T, 2010O, and 2018lrd; \citealt{li98circ,mattila10cbet,perley20circ}), two as type IIb (SN2005U and SN2010P; \citealt{leonard05atel,ryder10cbet}), and one as type Ib/c (SN2020fkb; \citealt{tomasella20atel}).
Similarly to {\ngc}, Arp\,299 has a high SN Ib/c rate compared to the SN II rate \citep{anderson11}, which is common for hosts with multiple SNe \citep{anderson13}. 
The high SN rate of Arp\,299 is consistent with its high SFR of $\sim70\,\msunyr$ \citep{thone09}.

This paper has three objectives. The first is to determine whether the SFR of {\ngc} can explain the high SN rate. The second is to test whether SNe in {\ngc} were born in concentrations of atomic gas indicating a recent gas accretion. The third is to test whether the environment of {\ngc} might have been responsible for its high SN rate.

We adopted a redshift for {\ngc} of $z=0.00649$ \citep{haynes97}, a luminosity distance of 27.9\,Mpc, and a corresponding scale of 0.134\,kpc arcsec$^{-1}$.
This  assumes a cosmological model with $H_0=70$ km s$^{-1}$ Mpc$^{-1}$,  $\Omega_\Lambda=0.7$, and $\Omega_{\rm m}=0.3$.

\section{Selection and data}
\label{sec:data}

\begin{table*}
\caption{{\hi} fluxes and atomic gas masses of galaxies in the {\ngc} group.}
\centering
\begin{tabular}{lcccccccccc}
\hline\hline
Galaxy & RA / deg & Dec / deg & Dist / $\arcmin$ & Dist / kpc & \multicolumn{3}{c}{$\mbox{Flux} / \jykms$} & \multicolumn{3}{c}{$\mhi / 10^9\msun$} \\
 & & & & & 18\arcsec & 33\arcsec & 60\arcsec &  18\arcsec & 33\arcsec & 60\arcsec \\
\hline
NGC2770 & 137.3904540 & 33.1235300 & $\cdots$ & $\cdots$  & 38.9 & 38.6 & 34.8 & 7.14 & 7.08 & 6.40 \\
NGC2770-NW & $\cdots$ & $\cdots$  & $\cdots$ & $\cdots$  & 23.0 & 22.3 & 20.2 & 4.22 & 4.10 & 3.72 \\
NGC2770-SE & $\cdots$ & $\cdots$  & $\cdots$ & $\cdots$  & 17.5 & 17.6 & 16.1 & 3.21 & 3.22 & 2.96 \\
NGC2770B & 137.3325000 & 33.1205560 & 2.9 & 23 & 2.3 & 2.2 & 1.8 & 0.43 & 0.41 & 0.34 \\
NGC2770C & 137.4448689 & 33.1438283 & 3.0 & 24 & 0.9 & 0.8 & 0.6 & 0.17 & 0.15 & 0.11 \\
NGC2770D & 137.5071249 & 32.9498569 & 12.0 & 96 & 3.4 & 3.4 & 2.7 & 0.63 & 0.62 & 0.49 \\
NGC2770E & 137.6322999 & 33.0583291 & 12.8 & 103 & $<$0.3 & $<$0.3 & $<$0.4 & $<$0.05 & $<$0.05 & $<$0.08 \\
\hline
\end{tabular}
\label{tab:comp}
\tablefoot{Fluxes and masses are shown for three datasets at resolutions indicated in the header. The first row corresponds to an elliptical aperture encompassing the entire {\ngc}, whereas the following two rows correspond to rectangle apertures encompassing the north-western and south-eastern halves. The fourth and fifth columns correspond to the projected distance from {\ngc}.}
\end{table*}

The SNe in {\ngc} were selected as part of a larger study of atomic gas in SN hosts (Gotkiewicz \& Micha\l owski, in prep.). We investigated all known SNe up to August 2018 with redshifts $z<0.1$ from the Open Supernova Catalog
and searched the NASA/IPAC Extragalactic Database (NED) for {\hi} data for their hosts. The study of SNe in M74 was published in \citet{michalowski20}. 
The archival {\hi} data for {\ngc} are from Westerbork observations of neutral Hydrogen
in Irregular and SPiral galaxies (WHISP; \citealt{swaters02}), taken with the Westerbork Synthesis Radio Telescope (WSRT). The data are shown in Figs.~\ref{fig:zoom} and \ref{fig:image}. The synthesised beam sizes are $24\arcsec\times13\arcsec$, 
$36\arcsec\times30\arcsec$, and $65\arcsec\times56\arcsec$.

\section{Results}
\label{sec:results}

Figure~\ref{fig:zoom} shows the distribution of atomic gas in {\ngc}. There is clearly significantly more gas in the north-western than in the south-eastern part of the galaxy. The velocity field is regular and reflects a rotating disc.

Three SNe exploded 
in the near or far sides of the galaxy, close to the line of sight towards the centre  (see Fig.~1 of \citealt{thone17}).
SN\,1999eh exploded in an atomic gas concentration visible on the  {\hi} image with the highest resolution. SN\,2007uy and 2015bh are $\sim15\arcsec$ (2\,kpc) from the atomic gas cloud, south of the galaxy centre. 
However, projection effects make it impossible to securely associate SNe with these structures.
On the other hand, SN\,2008D exploded in the  north-western part of the galaxy, which has more atomic gas. Its position is $\sim30\arcsec$ (4\,kpc) from the centre of the region with the highest {\hi} surface density.

In Fig.~\ref{fig:image} we show the large-scale environment of {\ngc}.
The main atomic gas disc is embedded in faint emission extending 5{\arcmin} (40\,kpc) from the galaxy centre.

A companion galaxy called {\ngc}B in \citet{fynbo08gcn}, \citet{soderberg08}, and \citet{thone09} is detected in {\hi} $\sim3\arcmin$ (24\,kpc) to the west from the centre of {\ngc}.
This galaxy was first reported by \citet{garciaruiz02}, as KUG\,0906+333A, based on the WHISP data we are using.
Its velocity is clearly inconsistent with the rotation of the main disc of {\ngc}, having a negative velocity with respect to the systemic velocity, whereas the nearby north-western tip of the galaxy has a positive velocity.  {\ngc}B is connected to the north-western tip of {\ngc} by a clear {\hi} bridge. A similar weaker feature is present on the other side of the galaxy extending east from the south-eastern tip of the galaxy.
These features are labelled on Figs.~\ref{fig:zoom} and \ref{fig:image}.

To the east of {\ngc}, also $\sim3\arcmin$ (24\,kpc) away,  \citet{garciaruiz02} reported the existence of an {\hi} object. As they did not name it, we assigned a designation of {\ngc}C to it. It is marked on Figs.~\ref{fig:zoom} and \ref{fig:image}. In NED and the Sloan Digital Sky Survey (SDSS; \citealt{sdssI,sdssIV}) it is listed as \object{GALEXASC J090946.88+330840.4} and \object{SDSS J090946.76+330837.7}, respectively. Its detections in the ultraviolet and optical imply it is also a dwarf galaxy, not just a gas cloud. The SDSS does not list a redshift for this galaxy\footnote{\url{http://skyserver.sdss.org/DR16//en/tools/explore/summary.aspx?id=1237664869211505073}}.

\citet{garciaruiz02} also reported a companion galaxy $\sim12\arcmin$ (96\,kpc) to the south-east. We named this galaxy {\ngc}D.
It is listed in  SDSS as SDSS J091001.70+325659.4\footnote{
\url{http://skyserver.sdss.org/DR16//en/tools/explore/summary.aspx?id=1237661384379924626}} and has a spectroscopic redshift of $0.00674\pm0.00003$, consistent with the {\hi} redshift and only $75\,\kms$ from the redshift of {\ngc}.


Finally, we searched the SDSS\footnote{\url{http://skyserver.sdss.org/dr16/en/tools/search/sql.aspx}} for additional galaxies with redshifts $0.006<z<0.01$ within the WHISP dataset and found SDSS J091031.75+330329.9\footnote{\url{http://skyserver.sdss.org/DR16//en/tools/explore/summary.aspx?id=1237664869211570362}} at $z=0.00684829$. It is $107\,\kms$ and   $12.8\arcmin$ (103\,kpc; to the east) from {\ngc}. We named it {\ngc}E. It is not detected in the WHISP data.

Figure \ref{fig:opt} shows all these nearby companions on the SDSS DR16\footnote{\url{http://skyserver.sdss.org/dr16/}} image.
Table~\ref{tab:comp} shows the results of {\hi} aperture photometry for galaxies covered by the WHISP data. For {\ngc} we used an elliptical aperture with semiaxes of  $160$ and $100\arcsec$, with the angle between the semi-major axes and the north of 30 degree (towards the west). We also used two square apertures encompassing the north-western and south-eastern halves of the galaxy. For {\ngc}B, C, and D we used circular apertures with radii of $60\arcsec$. 
In order to estimate the {\hi} upper limit for {\ngc}E, we first calculated a typical root mean square of the cubes per nominal $20\,\kms$ channel of 1.9, 2.2, and 3.1\,mJy\,beam$^{-1}$ for 18, 33, and 60{\arcsec} resolutions, respectively. Then we assumed that this galaxy is a point source and that its {\hi} line spans $100\,\kms$, and we reported $3\sigma$ upper limits.

The north-western half has $25$--$30$\% more atomic gas than the south-eastern part.
The total {\hi} flux and atomic gas mass of {\ngc} are consistent with  those derived by \citet[][$34\,\jykms$ and $7.04\times10^9\,\msun$]{rhee96} and \citet[][$36.9\pm1.4\,\jykms$ and $9.76\times10^9\,\msun$]{matthews01}. The {\hi} masses of {\ngc}B and D are around 20 times lower than that of {\ngc}, whereas {\ngc}C has 40 times less atomic gas than {\ngc}. {\ngc}E has at least 100 times less atomic gas than {\ngc}.

We also investigated galaxies  further away from {\ngc} that are not covered by the WHISP data. 
Within 60{\arcmin} (482\,kpc) and $\pm500\,\kms$ of {\ngc} (velocity of $1945.65\,\kms$), NED lists four galaxies:
{\ngc}B, {\ngc}D, 
and WISEA J091312.36+332639.2\footnote{Or SDSS J091312.37+332640.2 at $z=0.00607566$; \url{http://skyserver.sdss.org/DR16//en/tools/explore/summary.aspx?id=1237664869748637939}} 50{\arcmin} (400\,kpc) to the north-east at a redshift of 0.006047 ($1813\,\kms$),
and WISEA J090706.29+322219.5\footnote{Or SDSS J090706.26+322219.4 at $z=0.006475154$; \url{http://skyserver.sdss.org/DR16//en/tools/explore/summary.aspx?id=1237661126154125569}}, 55{\arcmin} (440\,kpc) to the south-west at a redshift of 0.005814
 ($1743\,\kms$). 
The SDSS does not list any additional galaxies within these search criteria.
Moreover, \citet[][their Table~7]{garciaruiz02} reported three other galaxies within 100{\arcmin} (800\,kpc) of {\ngc}: IC 2445 (91.4{\arcmin}, 735\,kpc),
CG 0010 (96.1{\arcmin}, 773\,kpc), and
UGC 04777 (96.7{\arcmin}, 777\,kpc).
{\ngc} is not covered by the Arecibo Legacy Fast ALFA Survey (ALFALFA; \citealt{haynes18}), prohibiting a systematic {\hi} search of companions.


\section{Discussion}
\label{sec:discussion}

\subsection{Enhanced SFR explaning high SN rate}

We now present the evidence that the SFR of {\ngc} has been recently 
enhanced, which explains the high SN rate.
The baseline non-enhanced SFR can be calculated in several ways. The main-sequence SFR for a stellar mass of $2\times10^{10}\,\msun$ \citep{thone09} at
$z=0.00649$ is $1.1\,\msunyr$  \citep{speagle14}. The atomic gas mass $\mhi=7.14\times10^9\,\msun$ (Table~\ref{tab:comp}) predicts $\mbox{SFR}=1.9\,\msunyr$ \citep[][Eq.~1]{michalowski15hi}.  Finally, the total molecular gas mass of $\log(\mhtwo/\msun)=8.9$ \citep{vanderhorst11} predicts  $\mbox{SFR}=0.8\,\msunyr$ \citep[][Eq.~1]{michalowski18co}.

On the other hand, the level of the expected enhanced SFR can be estimated from the measured SN rate ($R_{SN}$). The scaling of \citet{strolger15} of $R_{SN}\,(\mbox{SN\,yr}^{-1})=0.01\times\mbox{SFR}\,(\msunyr)$ as well as four SNe in 20 years imply $\mbox{SFR}=20\,\msunyr$.
This calibration takes into account all SNe (i.e.~progenitors masses larger than $8\,\msun$). {\ngc} hosted SNe type Ib and IIn, so, if their progenitors were more massive, the SFR required to explain their frequency would be higher. On the other hand, a double-star channel progenitor predicts lower masses \citep{podsiadlowski92}, consistent with the range analysed by \citet{strolger15}.

The SFR of {\ngc} measured from continuum estimators not affected by dust attenuation (infrared, radio, and full spectral energy distribution modelling) is $\mbox{SFR}=1.1$--$1.2\,\msunyr$ \citep{thone09}.
On the other hand, the dust-uncorrected ultraviolet SFR was measured to be $0.5\,\msunyr$ \citep{thone09}. The comparison with the radio estimate suggests a low level of dust attenuation of $A_{\rm UV} = 2.5 \log(\mbox{SFR}_{\rm radio}/\mbox{SFR}_{\rm UV})=0.95$\,mag \citep{michalowski12mass} or $A_V = A_{\rm UV}/2.2=0.43$\,mag,  assuming a Small Magellanic Cloud extinction curve \citep{gordon03}.
These low values do not imply that the SFR is enhanced.
However, if 
the enhancement
is recent (of the order of 10\,Myr), then these SFR estimates are not expected to increase because these indicators trace timescales of the order of 100\,Myr. In the case of a recent enhancement, only the  SFRs from line indicators (H$\alpha$, [OII], [CII]) are expected to increase because they trace much shorter timescales of the order of 10\,Myr. 

Based on the H$\alpha$ narrowband filter observations, \citet{thone09} reported a total SFR of $0.42\,\msunyr$ without dust correction. The median and mean colour excess measured in regions of {\ngc} are $E(B-V)=1.64\pm0.24$ and $1.67\pm0.15$\,mag, respectively \citep{thone09}.
This attenuation level is much higher than that experienced by the continuum emission.
We assumed the slope of the attenuation curve $R_V\equiv A_V/E(B-V)=3.1$ and that the extinction at the V-band ($A_V$) is similar to that at the wavelength of the H$\alpha$ line \citep{gordon03,michalowski12grb}. This corresponds to dust-corrected H$\alpha$ SFR of $45.4\pm0.8$ and $49.1\pm0.7\,\msunyr$. 

These high values indicate a significant enhancement in star formation activity, and they can fully explain the high SN rate in {\ngc}.
The merging system Arp\,299 does not resemble the morphology of {\ngc}, but it has a similarly high SFR \citep{thone09}, and its high SN Ib/c rate was also claimed to be a result of the  young age of the most recent star formation episode \citep{anderson11}.

The SFR enhancement in {\ngc} must have happened recently (of the order of 10 Myr ago) because the SFRs based on the infrared and radio have not yet had time to increase.
High line-to-continuum SFR ratios were also reported for gamma-ray burst (GRB) hosts and interpreted as recently enhanced star formation activity  \citep{michalowski16}.

Hence, the unique feature of {\ngc} compared to other galaxies is the fact that it was observed very shortly after the enhancement of the SFR. It also has high dust extinction: $E(B-V)$ above 1 mag is very rare and is found only in galaxies with SFRs of a few tens of $\msunyr$ \citep{kewley06,wijesinghe11,kreckel13,xiao12,catalantorrecilla15,davies16,wang16}.

Other host galaxies of multiple SNe type Ib/c also have sufficiently high SFRs to explain the measured SN rate.
In the sample of galaxies having hosted many SNe compiled by \citet{thone09}, apart from {\ngc}, there are six other galaxies with two SNe type Ib/c\footnote{\object{NGC 3690}, \object{NGC 2207}, \object{NGC 3810}, \object{NGC 7714}, \object{NGC 4568}, \object{NGC 3464}.}. Out of the five with measured radio SFRs, three have high values of $6$--$70\,\msunyr$. NGC\,3810 and NGC\,7714 have radio SFRs around $1\,\msunyr$, but their SFRs measured with the H$\alpha$ line are a few times higher, $4$--$6\,\msunyr$ \citep{james04}. That implies that their SFRs have been recently enhanced, like that of {\ngc}.
Two SNe in 20 years would imply an expected SFR of $10\,\msunyr$ (as above from the calibration of \citealt{strolger15}). Hence, within likely errors in this calibration, these SFRs are consistent with the measured SN rate.
We note that out of the six galaxies that hosted more than one SN Ib/c, only two (NGC\,3464 and 3810) have undisturbed spiral morphology similar to that of {\ngc}. The remaining four are strongly interacting systems with strong disturbances.

\subsection{Interaction resulting in an enhanced SFR}

The {\hi} bridge between {\ngc} and {\ngc}B provides definitive proof that they are interacting. This was  suggested by \citet{fynbo08gcn}, \citet{soderberg08}, and \citet{thone09}, but they did not find any obvious perturbations in the optical images of {\ngc}. This highlights the importance of {\hi} observations in looking for interaction signatures. The bridge, together with the feature on the opposite side of {\ngc} (at the south-eastern tip of the galaxy), form a characteristic tidal feature that is always seen in merger simulations  \citep{hopkins06b,hayward12,hayward14,pettitt16,oh16}.  The {\hi} asymmetry of the main disc in {\ngc} is also consistent with this scenario.
There is also more molecular gas in the north-western part of the galaxy \citep{vanderhorst11}, so the {\hi} deficit in the south-eastern part is not due to more rapid conversion to the molecular phase.
{\ngc} may also be interacting with {\ngc}C, D, and E, but this needs to be tested with deeper {\hi} observations.
The interaction with {\ngc}C is especially likely given its proximity to {\ngc}.

The interaction with these galaxies might have led to enhanced star formation activity, which would explain the high SN rate in {\ngc}. This is consistent with the study of galaxies hosting several SNe implying that a significant fraction of those hosting SN type Ib/c are classified as starbursts, irregular, or interacting, unlike those hosting SNe type II \citep{thone09}.
Similarly, in the sample of SN hosts analysed by \citet{galbany18}, SNe type Ib/c are associated with regions with the highest SFRs (their Figs.~5 and 9).
Moreover, \citet{arabsalmani15b,arabsalmani19} provided evidence of an on-going interaction for a GRB host.
This is likely because progenitors of SN type Ib/c and GRBs are more massive stars, so they are found preferentially in galaxies with recent enhancement of star formation. 

In the previous section we showed that in order to explain the discrepancy between the SFRs based on continuum and line estimators, a recent enhancement of the SFR in {\ngc} is needed. Indeed,  simulations of interacting galaxies show that SFRs rise on the timescale of a few tens of Myr or quicker \citep{hayward14,fensch17}.

\subsection{Environment of the SNe}

The scenario of a recently enhanced SFR implies that the SNe in {\ngc} are born in recently formed dusty star-forming regions. Indeed, these SNe were reported to be significantly affected by dust.
\citet{roy13} reported  $E(B-V)=0.63$\,mag for SN\,2007uy;
\citet{malesani09} reported $E(B-V) = 0.8$\,mag for SN\,2008D and \citet{thone17} measured high pre-explosion Balmer decrement for SN\,20015bh, which implies $E(B-V)=0.93$\,mag.

It is unclear if the SNe in {\ngc} are associated with high  atomic gas surface density regions. They appear to be close to peaks of {\hi} distributions, but only SN\,1999eh exploded within such concentration. Moreover, SN\,2008D exploded close to the highest {\hi} peak in the galaxy.

Hence, type Ib SNe have different environments than GRBs and broad-lined SNe type Ic (IcBL), which were shown to happen close to significant asymmetric off-centre {\hi} concentrations \citep{michalowski15hi,michalowski16,michalowski18,michalowski20,arabsalmani15b,arabsalmani19}. This suggests that the progenitors of type Ib SNe are not born out of recently accreted atomic gas, as was suggested for GRB and IcBL SN progenitors\footnote{Only AT\,2018cow, with an unclear nature, did not happen in an off-centre asymmetric {\hi} feature  \citep{michalowski19} and possibly only close to a gas ring \citep{roychowdhury19}.}. 

\subsection{Rotation}

Van der Horst et al.~(2011)
concluded that the rotation of the CO disc (high velocities in the north-western part; see their Fig.~4) is opposite to the {\hi} rotation. They referred to the position-velocity diagram of \citet[][their Fig.~4]{rhee96}, which shows low velocities in the north-western part. However, we believe that the galaxy sides are mis-labelled on Fig.~4 of \citet{rhee96} because the WHISP data show high velocities in the north-western part (Fig.~\ref{fig:zoom}), consistent with the CO rotation. Moreover, Fig.~4 of \citet{rhee96} shows that the galaxy side with more atomic gas (labelled SE there) has high velocities, so it should correspond to the north-western part. Hence, the conclusion of \citet{vanderhorst11} that the CO emission in {\ngc} is a result of a recent merger with a dwarf is likely incorrect. {\ngc} is currently interacting with its companion, but there is no evidence for a recent merger.   
We also note that the counter-rotating component at low velocities (2{\arcmin} north at $-100\,\kms$), which \citet{vanderhorst11} noticed in the data of \citet{rhee96}, is {\ngc}B.

\section{Conclusions}
\label{sec:conclusion}

{\ngc} has hosted three type Ib  and one type IIn SNe over the last 20 years. This high SN rate has not been fully explained and is not compatible with the previously reported SFR. 
We used archival {\hi} line data for {\ngc}
and reinterpreted the H$\alpha$ and optical continuum data. 

Even though the continuum-based SFR indicators do not yield high values, the dust-corrected H$\alpha$ luminosity implies a high SFR, consistent with the high SN rate. Such a disparity between the SFR estimators is an indication of recently enhanced star formation activity because the continuum indicators trace long timescales of the order of 100\,Myr, unlike the line indicators, which trace timescales of the order of 10\,Myr.

Hence, the unique feature of {\ngc} compared to other galaxies is the fact that it was observed very shortly after the enhancement of the SFR. It also has high dust extinction, $E(B-V)$ above 1 mag.

We provide support for the hypothesis that the increased SFR in {\ngc} is due to the interaction with its companion galaxies. The {\hi} bridge between {\ngc} and {\ngc}B is definitive proof that they are interacting. The atomic disc of {\ngc} is also asymmetric. The interaction is not obvious on the optical image, which does not reveal disturbances.  
We report the existence of a total of four companions within 100\,kpc  (one identified for the first time).

Supernovae  in {\ngc} exploded close to regions rich in atomic gas, but there are no {\hi} concentrations as dramatic as those detected for hosts of GRBs and type IcBL SNe.
This suggests that the progenitors of type Ib SNe are not born out of recently accreted atomic gas, as was suggested for GRB and IcBL SN progenitors.

\begin{acknowledgements}

We wish to thank 
the referee, Artur Hakobyan, for helpful suggestions and for pointing out the existence of {\ngc}C and E,  and
Joanna Baradziej for discussion and comments.

M.J.M.~acknowledges the support of 
the National Science Centre, Poland through the SONATA BIS grant 2018/30/E/ST9/00208, and of the Polish-U.S. Fulbright Commission.
A.d.U.P.~and C.C.T.~acknowledge the support from Ram\'on y Cajal fellowships (RyC-2012-09975 and RyC-2012-09984) and from the Spanish research project AYA2017-89384-P.
A.L.~acknowledges the support of 
the National Science Centre, Poland through the SONATA BIS grant 2018/30/E/ST9/00208.
J.H.~was supported by a VILLUM FONDEN Investigator grant (project number 16599).
M.P.K.~acknowledges support from the First TEAM grant of the Foundation for Polish Science No. POIR.04.04.00-00-5D21/18- 00.
P.K.~is partially supported by the BMBF project 05A17PC2 for D-MeerKAT.
The WSRT is operated by the Netherlands
Foundation for Research in Astronomy with financial support from
the Netherlands Organization for Scientific Research (NWO). 
Funding for the SDSS and SDSS-II has been provided by the Alfred P. Sloan Foundation, the Participating Institutions, the National Science Foundation, the U.S. Department of Energy, the National Aeronautics and Space Administration, the Japanese Monbukagakusho, the Max Planck Society, and the Higher Education Funding Council for England. The SDSS Web Site is http://www.sdss.org/.

The SDSS is managed by the Astrophysical Research Consortium for the Participating Institutions. The Participating Institutions are the American Museum of Natural History, Astrophysical Institute Potsdam, University of Basel, University of Cambridge, Case Western Reserve University, University of Chicago, Drexel University, Fermilab, the Institute for Advanced Study, the Japan Participation Group, Johns Hopkins University, the Joint Institute for Nuclear Astrophysics, the Kavli Institute for Particle Astrophysics and Cosmology, the Korean Scientist Group, the Chinese Academy of Sciences (LAMOST), Los Alamos National Laboratory, the Max-Planck-Institute for Astronomy (MPIA), the Max-Planck-Institute for Astrophysics (MPA), New Mexico State University, Ohio State University, University of Pittsburgh, University of Portsmouth, Princeton University, the United States Naval Observatory, and the University of Washington. 

We acknowledge the usage of the HyperLeda database (\url{http://leda.univ-lyon1.fr}).
This research has made use of 
the Open Supernova Catalog (\url{https://sne.space});
NASA/IPAC Extragalactic Database (NED) which is operated by the Jet Propulsion Laboratory, California Institute of Technology, under contract with the National Aeronautics and Space Administration;
SAOImage DS9, developed by Smithsonian Astrophysical Observatory \citep{ds9};
Edward Wright cosmology calculator \citep{wrightcalc};
the WebPlotDigitizer of Ankit Rohatgi ({\tt arohatgi.info/WebPlotDigitizer})
and NASA's Astrophysics Data System Bibliographic Services.
\end{acknowledgements}




\end{document}